\documentclass{aa}  
\usepackage{graphicx}
\usepackage{pdfpages}
\usepackage{subcaption}
\usepackage{float}
\usepackage{xcolor,colortbl}
\usepackage{natbib}
\usepackage{booktabs}
\usepackage{blindtext}

\bibpunct{(}{)}{;}{a}{}{,} 

\usepackage{txfonts}
\usepackage[colorlinks]{hyperref}
\begin{document}

   \title{The incidence of the LBV variability in the LMC}

   \author{V. M. Kalari,
          \inst{1}
          J. S. Vink\inst{2},
        C. Furey\inst{2,3},
                R. Salinas\inst{4},
A. Udalski
 \inst{5}
 M. Pawlak
 \inst{6}
        }

   \institute{Gemini Observatory/NSF’s NOIRLab, Casilla 603, La Serena, Chile\\
              \email{venu.kalari@noirlab.edu}
       \and
             Armagh Observatory and Planetarium, College Hill, BT61\,9DG Armagh, Northern Ireland, UK
        \and
        School of Mathematics and Physics, Queen’s University Belfast, Belfast BT7\,1NN, UK
        \and 
        Nicolaus Copernicus Astronomical Center, Polish Academy of Sciences, Bartycka 18, 00-716 Warszawa, Poland
        \and
        Astronomical Observatory, University of Warsaw, Al. Ujazdowskie 4, 00-478 Warszawa, Poland
        \and 
        Lund Observatory, Division of Astrophysics, Department of Physics, Lund University, Box 43, SE-221 00, Lund, Sweden
             }

   \date{Received  ; accepted  }

  \abstract
     {Luminous blue variables (LBVs) exhibit unique variability features, characterized by episodic outbursts ($>$1\,mag) accompanied by spectroscopic changes (S\,Dor variables).
     It is debated if all massive stars undergo an LBV-like phase during their evolution, or instead LBVs are exotic phenomena.} {We aim to quantify the incidence of LBV-like variability in the blue supergiant (BSgs) population of the Large Magellanic Cloud (LMC) using the OGLE survey.} {Here, we extend previous work in the Small Magellanic Cloud to the LMC, where we examine the light curves of 87 B Supergiants (BSgs) (out of 254 known BSgs) spanning timescales of twenty years, and 37 objects across a three year timescale for aperiodic variations resembling known S\,Dor variables.}{One blue supergiant, [ST92]\,4-13 shows S Dor type photometric variations. New spectra of this object reveals a potential change in spectral type compared to the literature classification. However, based on its spectral characteristics and low luminosity and mass, we do not currently classify it as an LBV.} {Our study highlights the need to classify bona fide LBVs as stars undergoing both photometric and spectroscopic variations. Based on currently known stellar population of S\,Dor variables in the LMC, the lifetime of the S\,Dor phase is at most $\sim$10$^3$\,yrs, in agreement with our duty cycle study based on OGLE data in the SMC. This is orders of magnitude shorter than assumed in literature. Our discovery of LBV-like variability at low luminosities may suggest that S\,Dor variations could arise from Eddington limit related physics over a wide range of stellar masses, rather than being linked to a unique evolutionary stage.}

  \authorrunning{Kalari et al.}

   \maketitle
 %

\section{Introduction}

Luminous blue variables (LBVs) are characterized by light curves with aperiodic decadal variations around 0.5--2 mag,  interspersed by micro-variations on the order of 0.1--0.2 mag (also termed S-Dor variables; \citealt{2001A&A...366..508V}), and rarely, dramatic outbursts \citep{conti}. The increase in brightness is accompanied by a color and spectral change, suggestive of an inflated envelope and a late spectral type in an eruptive phase, and a early type supergiant when quiescent. LBVs are rather rare, with the total number of known LBVs $\sim$20 across the Milky Way, and Magellanic Clouds. Currently, just 7 bona fide LBVs are currently known in the 1/2\,$Z_{\odot}$ Large Magellanic Cloud (LMC), with another 4 candidates, while only one bona fide member, R\,40, is known in the 1/5\,$Z_{\odot}$ Small Magellanic Cloud \citep{weiss}.

There are two potential evolutionary states for LBVs. In the standard picture of massive star evolution, they would be in a transitionary phase between the core hydrogen burning O star phase and the final Wolf-Rayet (WR) phase. This is oftentimes referred to as the Conti-scenario \citep{conti}. In this case, the LBV duration is usually considered to be relatively short (of order 10$^4$ years) in comparison to the bulk of the core He burning WR phase which would last several 10$^5$ years. This would be in rough agreement with the factor that there are hundreds of Wolf-Rayets in the LMC in comparison to just half a dozen of LBVs. 

A second possibility is however that LBVs do not make it all the way to the Wolf-Rayet phase, as they have been proposed as potential direct SN progenitors \citep{2006A&A...460L...5K, smith07, 2008A&A...483L..47T}. In this scenario, the LBV could either be much longer, say associated with the long core He burning lifetime of 10$^5$ years, and only a subset of LBVs would be in a pre-SN state, or alternatively, LBVs are really just pre-SN, and very rare, with a very short lifetime of say 10$^2-10^3$ years. In the latter scenario it has been proposed that LBV SNe could be the result from stellar merging \citep{2014ApJ...796..121J, 2015AAS...22534317T}.

Alternatively, \cite{2021A&A...647A..99G} suggest that LBV-like variations are triggered by wind-envelope interaction when stars approach the Eddington limit, as the mass loss rate increases with decreasing temperature that drives the inflation of the envelope, leading to a feedback loop corresponding to the observed variations. This model explains the observations as a physical phenomena, rather than being linked to a single evolutionary phase.

\begin{figure}
\centering
\includegraphics[width=9cm]{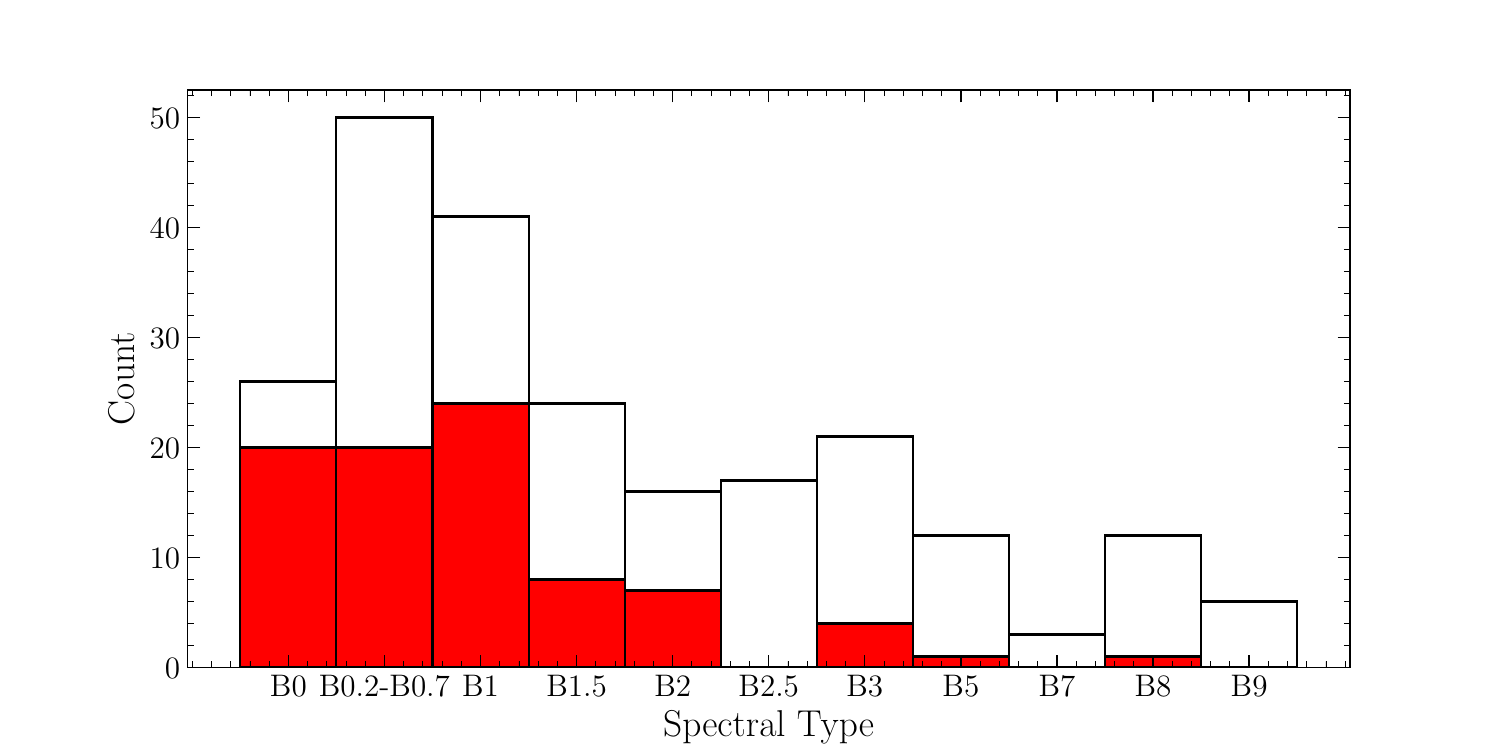}
\includegraphics[width=9cm]{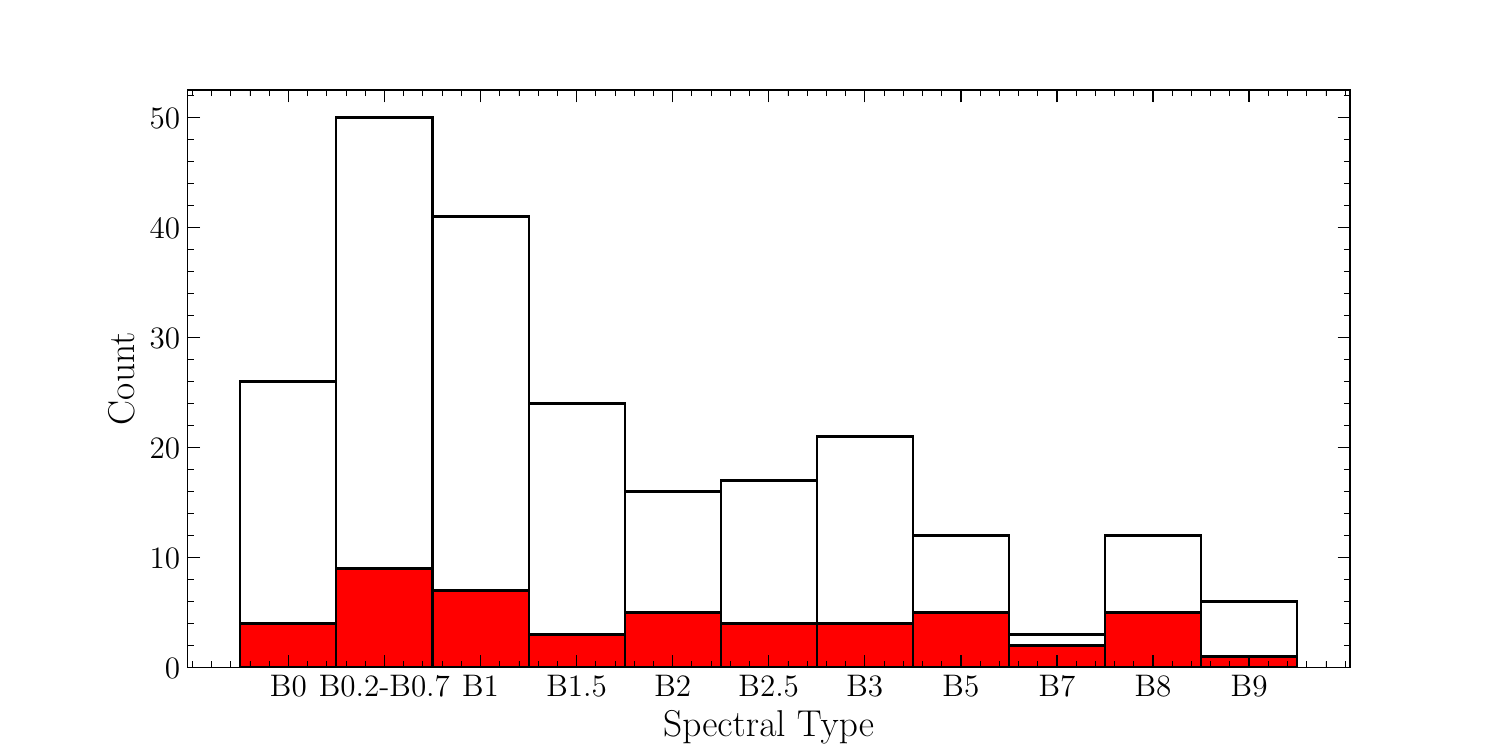} 
\caption{{\it Top}:Spectral types of Bsgs with OGLE data (red) and the total number of Bsgs from our sample (solid line). {\it Bottom}: The recovery fraction for {\it Gaia} multi-epoch photometry. Known LBVs are not shown. }
\label{sptype}
\end{figure}

\cite{2007AJ....134.2474M} studied the spectral similarity of known extra-galactic LBVs to generic extra-galactic supergiants, arguing the LBV phenomenon could be more widespread, perhaps by even or order of magnitude. On the other hand, from our SMC light-curve analysis \citep{Kalari18}, we found just 1 LBV candidate amongst the general population of B supergiants, suggesting the LBV phenomenon could be rare and special. However, a drawback of our SMC study was that it only considered 3 years of data, while the S Dor phase could easily last for decades \citep{2001A&A...366..508V}. This motivates our current study. In this paper, we expand the baseline for a similar number of objects in the LMC, but with 20 years of OGLE data, with nearly nightly cadence increasing the temporal coverage of objects nearly seven times.

Our paper is organized in the following manner. Section 2 presents the photometric data used in this study. The analysis of light curves used in this study is described in Section 3. Section 4 discusses the nature of two S\,Dor variables identified using the analysis. Section 5 discusses the incidence of S Dor variability among BSgs, and the light-curve analysis, with the conclusions outlined in Section 6.

\section{Data}

\subsection{Sample of Blue Supergiants}
We used the \cite{bonanos2009} catalog of massive stars in the LMC to construct our sample. \cite{bonanos2009} compiled spectral classifications for around 1750 massive stars in the LMC based on the literature. From this, we identified 254 BSgs and LBVs having traceable provenance. We excluded sources with uncertain or ambiguous spectral or luminosity classifications, but not binaries. While our choice of input catalog misses out on potentially cool phase LBVs, our main aim is not to discover new LBVs, but estimate the incidence of S Dor variability among known blue supergiants and thus does not affect our results. The resulting range of spectral subtypes is shown in Fig.\,\ref{sptype}. 

\begin{figure*}
\centering
\includegraphics[width=0.45\textwidth]{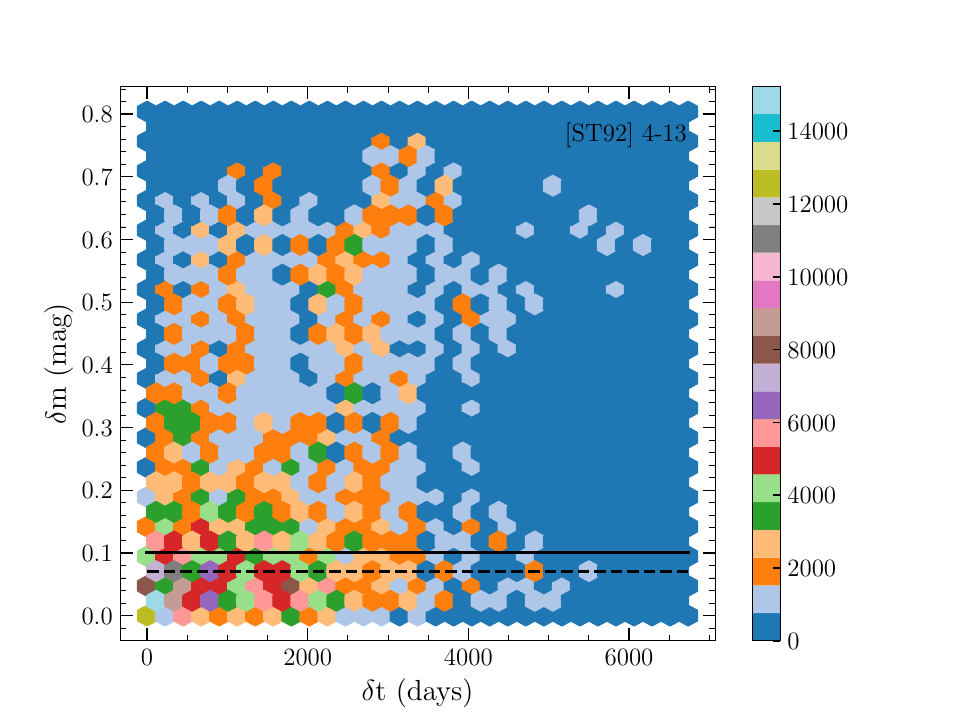} 
\includegraphics[width=0.45\textwidth]{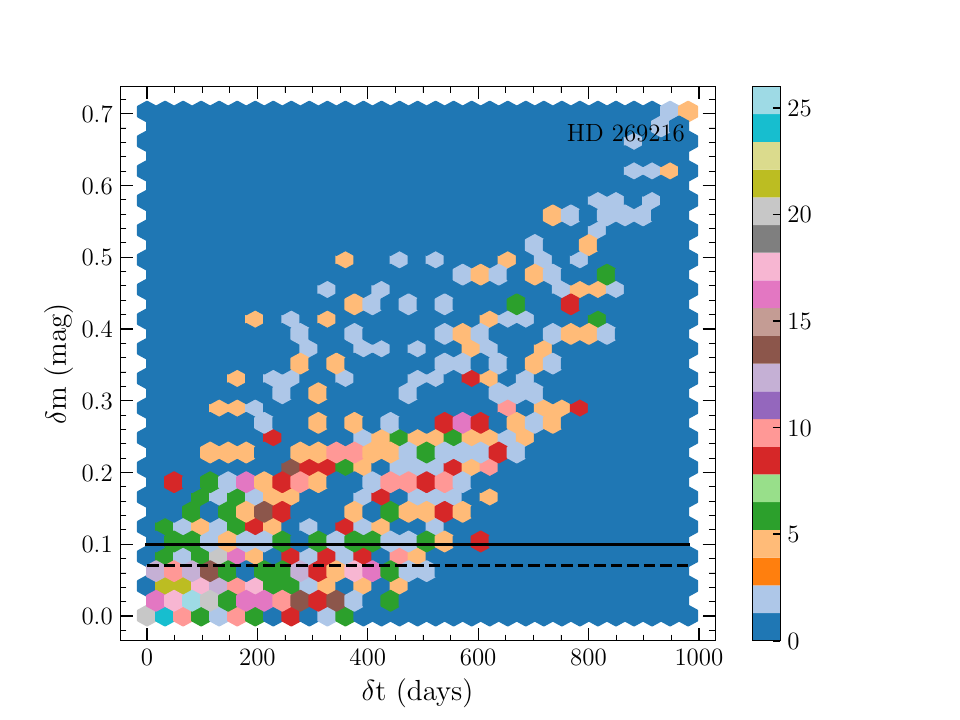} 
\caption{Representative example of the time series photometric analysis, where the color gradient marks the density of the points. The left panel shows the LBV candidate [ST92]\,4-13 $\delta$m-$\delta$t histogram computed using OGLE-IV data, with the $\delta$m levels of 0.08 (10$\sigma$) and 0.1\,mag marked. The right  panel shows the same for the known LBV HD\,269216, computed from {\it Gaia} multi-epoch photometry. }
\label{dmdt}
\end{figure*}

\subsection{OGLE multi-epoch photometry}
The Optical Gravitational Lensing Experiment (OGLE) survey provides high-cadence photometry in the Cousins $I$-band, with observations spanning approximately June 2001 to 2020, depending on the data release. The cadence varies from a few days to weekly sampling, depending on field and observing season, and the data are obtained using the 1.3\,m Warsaw telescope at Las Campanas Observatory in Chile. The median FWHM of the OGLE imaging is approximately $1\arcsec$, providing sufficient spatial resolution to resolve individual massive stars in the LMC. The reader is referred to \cite{ogleiv} and references therein for further details on the survey and data reduction.

We cross-matched these known BSgs with multi-epoch photometry from the OGLE-IV survey using a radius of 1$\arcsec$. In cases where OGLE-IV data were unavailable, we supplemented with OGLE-III data which provides shorter temporal coverage \citep{ogleiii}. A total of 83 unique matches were found in the OGLE-IV dataset, and 6 in the OGLE-III dataset. The OGLE dataset for each source contain typically around a thousand epochs per star, with coverage extending over nearly two decades. This enables the detection of both short and long-timescale variability phenomena. Photometric precision in the $I$-band is of the order 0.01\,mag for unsaturated sources, and systematics are minimized through the use of difference image analysis \citep{dia}.

\subsection{{\it Gaia} DR3 multi-epoch photometry}
{\it Gaia} DR3 includes multi-epoch photometry for selected bright sources \citep{eyer23}. This covers observations across the custom {\it Gaia} $G$, $BP$, and $RP$ filters spanning July 2014 to May 2017, covering between 30-80 epochs per target (except for Sk $-66^{\circ}$185, which has 175 epochs). Multi-epoch photometry were retrieved using the {\texttt{PYTHON astroquery}} package. Only photometric points satisfying the {\texttt{variability\_flag\_g\_reject}} (and in the respective filters for color analysis) were retained, and we ensured there are at least 25 epochs for source for analysis. In total, there are 49 BSgs with {\it Gaia} DR3 multi-epoch photometry. 11 of these also have OGLE IV light curves, and 1 has an OGLE III light curve. We also retrieved light curves of all confirmed LBVs (7 sources) from \cite{2018RNAAS...2..121R} which are not included in our input catalog of \cite{bonanos2009}. The analysis of these light curves are in Appendix A.

\subsection{Recovery and Completeness}

Figure~\ref{sptype} also shows the recovery fraction of BSgs in OGLE, and {\it Gaia} DR3 as a function of spectral subtype. We find that approximately 36\% of the BSgs identified by \citet{bonanos2009} have light curves in the OGLE datasets. For OGLE, this is mostly because of saturation, given the saturation limit around 12\,mag in $V$. This limit excludes all known LBVs in the LMC, which have apparent magnitudes brighter than $\sim$12\,mag in the optical \citep{2018RNAAS...2..121R}. Note that for B supergiants, on average, the magnitude stays relatively the same, or slightly increase with decreasing spectral class (for early Bs, from B0-B3 between 11--14 mag; and for latter types between 10--13 mag). A few objects fall into CCD gaps, and thus are not recovered. For the {\it Gaia} dataset, the recovery fraction is dependent on a variety of factors, and is expected to increase significantly in future releases \citep{eyer23}. 

The total fraction of stars with light curves is $\sim$49\%, and is biased towards earlier spectral types due to the OGLE dataset, however a few latter spectral types are recovered from the {\it Gaia} photometry. For this sample size, simple binomial properties can be estimated at the 99\% confidence level with a 8\% margin of error given the total sample size. We note that while our selection criteria are biased due to the saturation (and faintness) limit of the photometry, considering the B spectral class as a whole, the data does contain a sufficiently representative sample for such an analysis. The list of 126 unique BSgs analyzed and their properties are electronically available via CDS, along with the complete list of 254 BSgs that are shown in Fig.\,\ref{sptype}. An representative of this is shown in Table\,\ref{all}.

\begin{table*}
\centering
\caption{B supergiants in the Large Magellanic Cloud. }
\resizebox{\textwidth}{!}
{\begin{tabular}{llllllll}
\hline
  \multicolumn{1}{c}{Name} &
  \multicolumn{1}{c}{Right Ascension} &
  \multicolumn{1}{c}{Declination} &
  \multicolumn{1}{c}{Spectral type} &
  \multicolumn{1}{c}{Survey} &
  \multicolumn{1}{c}{Survey\,ID$^1$} &
  \multicolumn{1}{c}{Simbad Name} &
  \multicolumn{1}{c}{ID$^{2}$} \\
    \multicolumn{1}{c}{} &
  \multicolumn{1}{c}{(degrees)} &
  \multicolumn{1}{c}{(degrees)} &
  \multicolumn{1}{c}{} &
  \multicolumn{1}{c}{} &
  \multicolumn{1}{c}{} &
  \multicolumn{1}{c}{} &
  \multicolumn{1}{c}{} \\
\hline\hline
  BI\,103 & 78.2848 & $-$69.3011 & B2\,Ia & OGLE\,IV & LMC\,503.15\,93353 & 2MASS\,J05130832-6918041 & 388\\
 N11-017 & 74.0732 & $-$66.3051 & B2.5\,Iab & -- & -- & 2MASS\,J04561755-6618189 & 99\\
  NGC\,2004-007 & 83.0031 & $-$67.3396 & B8\,Ia & {\it Gaia} & DR3\,4660132807190729728 & Cl*\,NGC 2004\,ELS\,7 & 964\\
  Sk$-66$\,185 & 85.6271 & $-$66.3030 & B0\,Iab & OGLE\,IV & LMC\,555.29\,8272 & GSC\,08891-01385 & 1681\\
 &  &  &  & {\it Gaia} &  DR3\,4659925617961629824 &  & \\
  Sk$-$68\,23 & 75.2061 & $-$68.1194 & B3\,I & OGLE\,III & LMC\,125.2\,24072 & TYC\,9161-1217-1 & 276\\
\hline\end{tabular}}
 {\raggedleft Coordinates are given in the J2000 epoch. The table includes objects having multi-epoch OGLE or {\it Gaia} photometry. Five random rows are shown, the whole table consisting of 254 objects is available online. $^1$ Refers to either the OGLE ID when the object has OGLE data, or {\it Gaia} DR3 identifier when there is {\it Gaia} multi-epoch photometry. $^2$ Refers to the ID from Bonanos et al. (2009).}
 \label{all}
\end{table*}

\section{Analysis of time series data}

\subsection{$\delta m- \delta t$ histograms}

To examine aperiodic variability, we computed the $\delta$m-$\delta$t plot for each light curve as described in \cite{Kalari18}, which represents the incidence of variability at a particular timescale in a light curve. It is defined as the sum of all observations at magnitude $m$ and time $t$ of a light curve, by
computing the differences,
\begin{equation}
  \begin{array}{l}
\Delta m_{ij} = |m_{i}-m_{j}|~~ (i>j)\\
\Delta t_{ij} = |t_{i}-t_{j}|~~ (i>j)
  \end{array}
\end{equation} 

and $i>j$ such that each pair is only considered once. By presenting the differences between magnitude and time as a density histogram, we can directly observe how much variability is observed on what timescales. BSgs with $\delta$m greater than 0.1\,mag were examined visually. The resulting $\delta$m-$\delta$t plots for all sources are provided on request to the principal author. A representative example using OGLE and {\it Gaia} data are given in Fig.\,\ref{dmdt}.

From the $\delta$m-$\delta$t plots, we identify only three stars with OGLE data that show aperiodic variability above the significance threshold. The first is [ST92] 4-13 (alias W61\,4-1 in SIMBAD\footnote{\cite{bonanos2009} and SIMBAD list W61\,4-1 as B1\,Ia based on \cite{1998A&AS..130..527T}. However, this is likely an incorrect reference, as that paper does not study any stars in this cluster. The spectral type based on those coordinates appears to be based on the result of \cite{1992A&AS...92..729S}, where it is classified as star 4-13 in the cluster LH104, hence we adopt [ST92]\,4-13 as the reference name, see also \cite{skiff}}, and {\it Gaia} DR3\,4657649594536852352), a B1\,I star, with the magnitude changing by nearly 1\,mag over the past two decades. The other two are VFTS\,698, the known binary masquerading as a B[e] star \citep{dunstall12}, and VFTS\,652-- a known spectroscopic double lined binary \citep{howarth15}. We discard these two from our sample given that the magnitudes vary less than 0.5\,mag over the timescale, and only consider [ST92]4-13, as binarity is considered the likely origin of the magnitude variations in the latter two. 

\begin{figure}
\center
\includegraphics[width=0.4\textwidth]{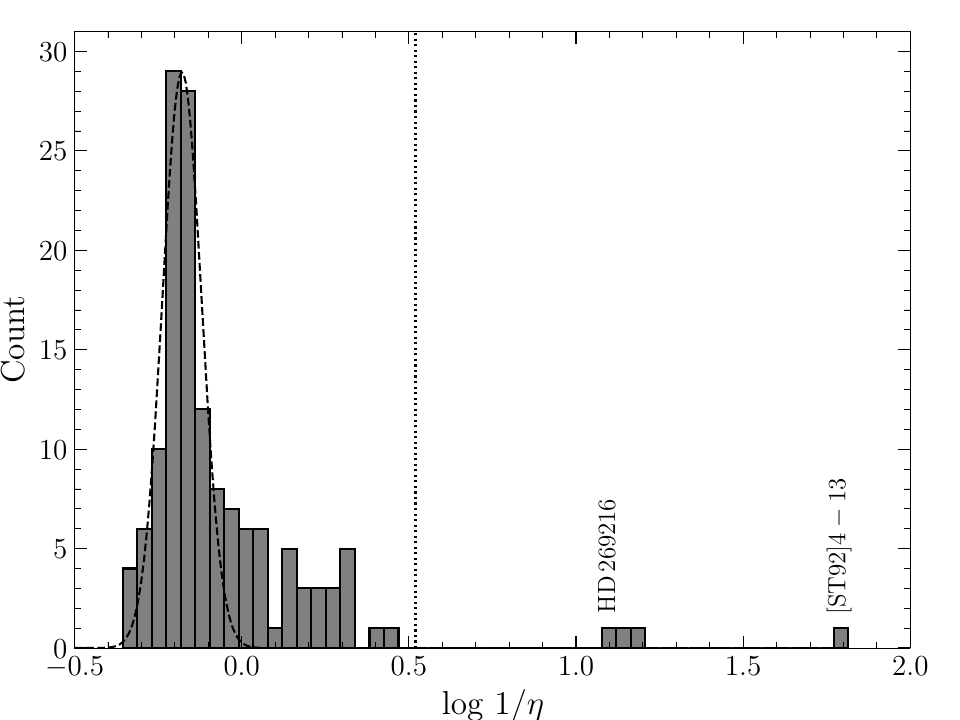} 
\caption{Histogram of 1/$\eta$ values computed using OGLE and {\it Gaia} light curves, in logarithmic scale. The positions of the identified LBVs, HD\,269216 and [ST92] 4-13 are marked. The best-fit Gaussian to the population except the outliers (log (1/$\eta) >$ 1) is shown by the dashed line. The dotted line shows the 3$\sigma$ cut-off.  }
\label{eta}
\end{figure}

\begin{figure*}[h!]
\center
\includegraphics[width=0.45\textwidth]{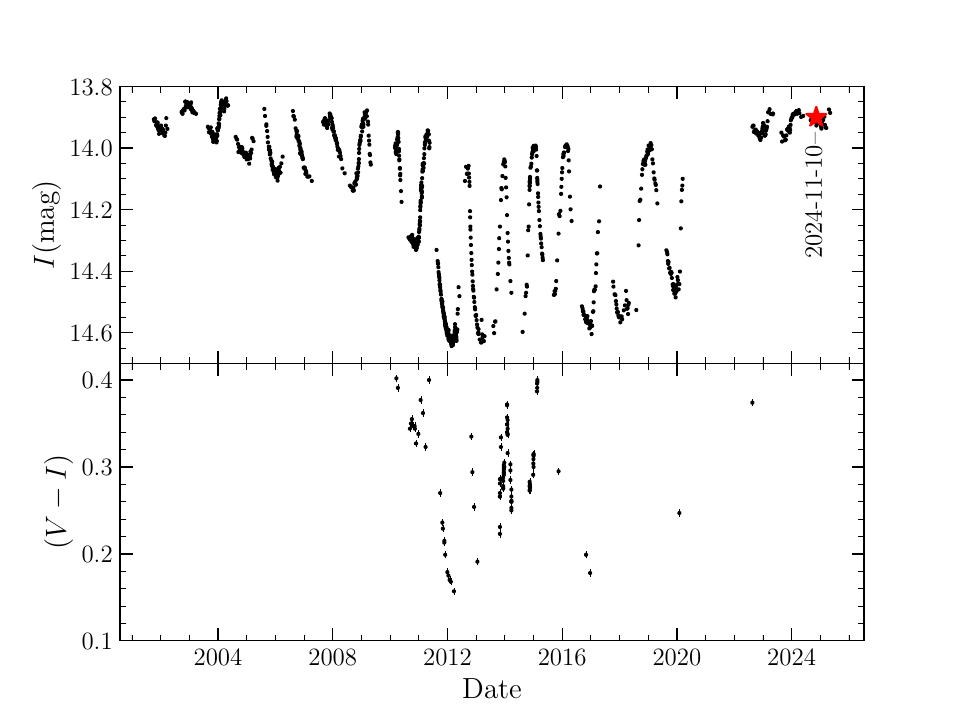}
\includegraphics[width=0.45\textwidth]{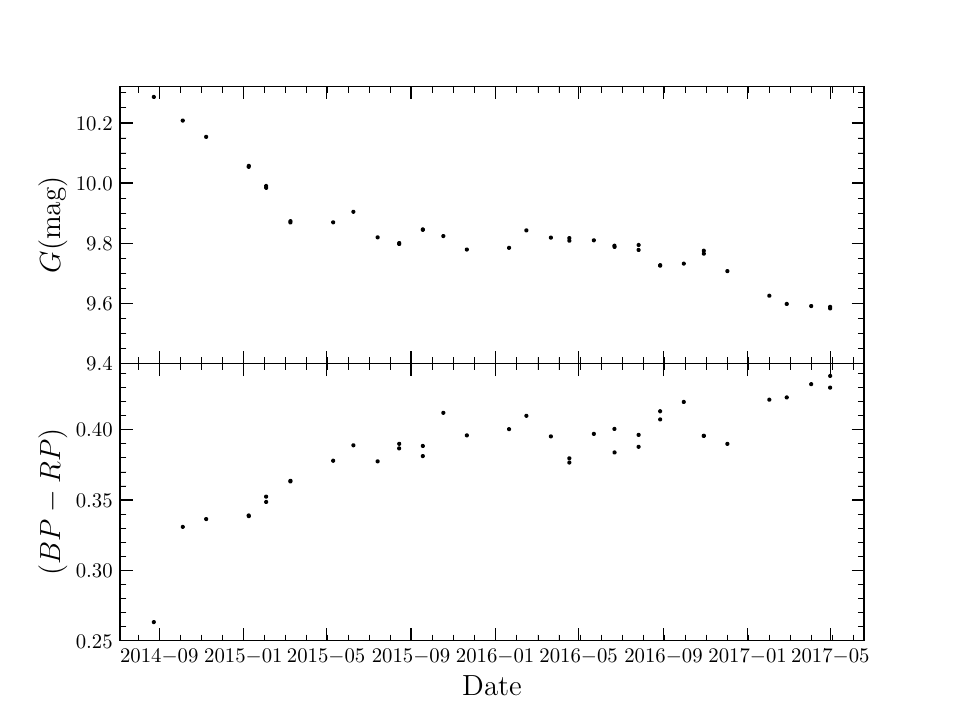} 
\caption{{\it Left}: $I$-band light curve of [ST92] 4-13, with the ($V-I$) color shown at the bottom. Data are from the OGLE-IV photometry. The red asterisk marks the epoch when the spectral data were taken. {\it Right}: {\it Gaia} $G$ and $(BP-RP$) magnitude and color light curves of HD 269216, shown at the top and bottom panel respectively.  }
\label{w641lc}
\end{figure*}

\begin{figure*}
\center
\includegraphics[width=1.05\textwidth]{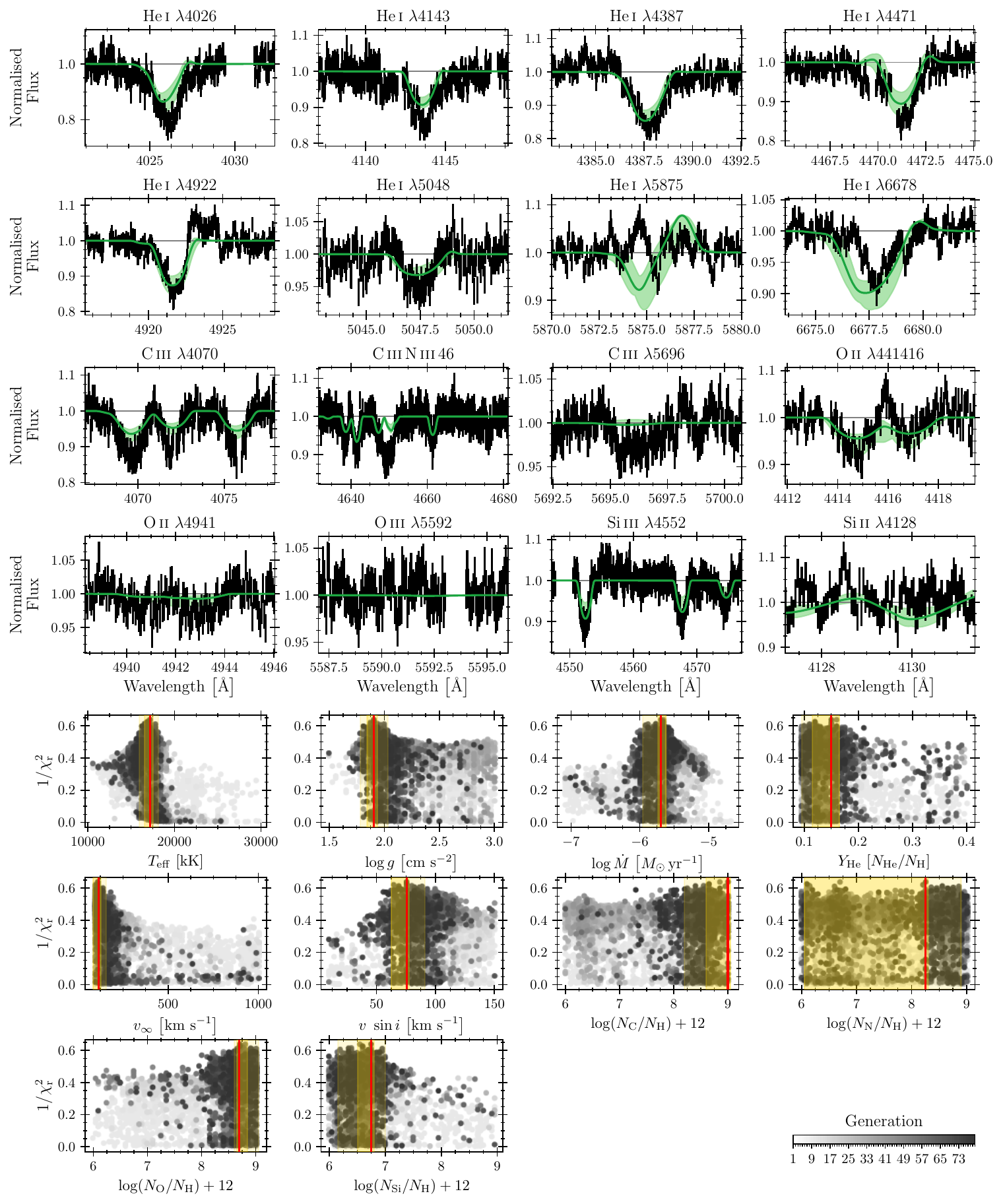} 
\caption{\textsc{Kiwi-GA} fit of [ST92] 4-13. See Section 4.1.2 for details. }
\label{gafit}
\end{figure*}

In the {\it Gaia} dataset, there is only one star with the $\delta m$-$\delta t$ histogram resembling a S\,Dor variable. HD\,269216 (HIP\, 24347, {\it Gaia} DR3\,4658204053297963392) is a known LBV, previously identified in \cite{prinja91} and \citet{massey17}. Eight other stars do show aperiodic variability above 0.1\,mag, but not a significant levels as seen in [ST92]4-13 or HD\,269216 (i.e. $<$\,0.5\,mag). This variability depends on a few ($<$5) epochs when examining the light curves, and given the scan-dependent possibility of spurious signals \citep{gaiaspur}, we refrain from examining these further without future epoch photometry confirming the variability.

\subsection{von Neumann $\eta$ index}

In addition to the $\delta m- \delta t$ histograms, a method to identify irregular variability in irregular spaced data is via the von Neumann index \citep{shin}. This index is computed as the mean squared successive difference over the variance of the distribution and acts as an indicator of independence for a series of observations \citep{von}. In photometric monitoring observations, it identifies whether the underlying distribution is smooth. We adopt the 1/\,$\eta$ value as defined in \citet{variability} as a variability indicator, where larger values indicate irregular variability.

Our analysis results are shown in Fig.\,\ref{eta}. Here, the histogram of 1/\,$\eta$ values for our sample is shown in logarithmic scale, computed for both the {\it Gaia} and OGLE light curves. Three stars in the OGLE sample have values 3$\sigma$ away from the mean of the Gaussian fitted to the distribution (peaking at $\mu$=$-$0.15 in log scale). These are [ST92]\,4-13 with a log(1/\,$\eta$) value of 1.81, LH\,104\,4-67 (1.13), and LH\,101\,5-18 (1.18). On examining the light curve of the latter two, we find that a jump between epochs in the mean magnitude, rather than any aperiodic behavior causes these high values. In the {\it Gaia} dataset, HD\,269216 shows a value discrepant from the median of 1.11. Overall, a comparison with the von Neumann index suggests that the same stars would have been identified for aperiodicity. The functions used to compute both the $\delta m-\delta t$ histograms, and the von Neumann $\eta$ index are publicly available{\footnote{https://github.com/astroquackers/aperiodic}}.

\section{The S\,Dor variables:  [ST92]\,4-13 and HD\,269216}

\subsection{ [ST92]\,4-13}

\subsubsection{Photometric variability}
Fig.\,\ref{w641lc} shows the light curve of the candidate S\,Dor variable--[ST92]\,4-13 spanning nearly two decades. The light curve displays oscillations on the timescales of few months, and the brightness decreased by 0.7\,mag from 13 June 2009 to May 2012, and then increased with a larger oscillation range. In a Lomb-Scargle periodogram, we found at least four significant peaks, three around a few days and a longer one at 7 years. Large amplitude variations have ceased since 2022. Fig.\,\ref{w641lc} visualizes the aperiodic change seen, showing a magnitude increase of nearly 0.8\,mag seen over timescales of a year. The nearly yearly period seen along with the large amplitude variations are consistent with the changes seen for S\,Dor variables \cite{2001A&A...366..508V}. For e.g., \cite{2001A&A...366..508V} report a period of 371.4 days for AG\,Car with beat cycles. {\it Gaia} indicators of multiplicity ({\texttt{RUWE} and {\texttt{ipd\_fmp}) that have recovered binaries in early emission line B-type stars \citep{kalari25}, do not indicate that there is an unseen companion that may cause the observed light curve. Based on this information, we classify [ST92]\,4-13 as a S\,Dor variable. Note that further unreleased OGLE observations taken between 2020-2024, which were not part of the original analysis were examined for [ST92]\,4-13 and are shown in Fig.\,\ref{w641lc}.

\subsubsection{Spectral analysis}
Investigating whether there might be spectral changes accompanying the aperiodic light curve seen, we search the literature for for spectra. \cite{1992A&AS...92..729S} note it as having ($B-V$) of $-$0.09 and ($U-B$) of $-$0.82 with data taken between 1988 and 1990, and a spectral type of B1\,Ia based on low resolution spectroscopic observations using the \ion{He}{i}$\lambda$4471/\ion{Mg}{ii}$\lambda$4481, and \ion{He}{i}$\lambda$4144/\ion{He}{i}$\lambda$4121 line ratios within the same time frame. The object is marked as an emission line star based on the catalog of \cite{howarthlmc}, however no line widths are provided. {\it Gaia} DR3 \citep{gaiadr3} XP mean spectra (reference epoch 2016.0) of the object shows a corrected H$\alpha$ line width of $-15\,\AA$. 

To observe if any changes in spectral type are seen, we obtained spectra using the Gemini high-resolution optical spectrograph (GHOST) at Gemini South \citep{ghost}, on November 10, 2024 (epoch marked in Fig.\,\ref{w641lc}; MJD of 60624.210). Data were taken in photometric conditions with seeing around 0.8$\arcsec$. Immediately after the science target we observed the spectrophotometric standard for CPD-69 177 for absolute flux calibration. The final spectra have $R\sim$ 47\,000, and reach signal-to-noise ratios around 50 near the $V$-band. 
The spectra display prominent Balmer emission lines, which consist of a broad and narrow emission component. The effective temperature is more representative of late-B type supergiant from the spectra, with change observed in the H$\alpha$ line width compared to the Gaia spectrum ($-20$\AA). The Balmer profiles do not exhibit the P\,Cygni morphology in the lower H lines seen in LBV candidates during quiescence (as shown in \citealt{2007AJ....134.2474M}) as predicted as the object is heading towards visual minimum in the light curve, but do show P\,Cygni profile in \ion{He}{i}. 

To estimate the stellar parameters of [ST92] 4-13 and get an idea of its evolutionary status, we used the genetic algorithm \textsc{Kiwi-GA} \citep{brands2022r136} in combination with the stellar atmosphere code \textsc{Fastwind} \citep[v10.6; ][]{santolaya1997atmospheric, puls2005atmospheric, rivero2012nitrogen, carneiro2016atmospheric, sundqvist2018atmospheric}. We describe the spectral reduction process and modeling approach in Appendix B. For a more in-depth explanation of the models, we refer readers to the cited literature. Briefly, \textsc{Kiwi-GA} is a Python wrapper for \textsc{Fastwind} that uses the concept of evolution to automatically fit \textsc{Fastwind} models to an observed spectrum. Over a predefined number of generations and number of individuals per generation, the desired parameter space is explored by means of reproduction and mutation to find, for a given input spectrum, best fitting stellar and wind parameters and statistical uncertainties. Here, we use 81 generations with 85 models per generation to estimate the stellar parameters of our target star.

The best-fitting parameters are given in Table \ref{tab:best_fit_params}, with the corresponding best-fitting spectrum and parameter posteriors shown in Fig.\,\ref{gafit}. The top four rows of the figure show the observed, continuum-normalized spectrum and error bars in black, the best fitting \textsc{Fastwind} model as the solid green line, and the 1$\sigma$ uncertainty region as the shaded green region. The bottom three rows show the posteriors for each free parameter. On each of these axes, the points represent the models of the \textsc{Kiwi-GA} run, whose colors are mapped to the generation number of the model. The red vertical line gives the best fitting value, and the dark and light yellow shaded regions give the 1- and 2$\sigma$ uncertainty regions, respectively. 

For [ST92] 4-13, we estimate an effective temperature $T_{\rm eff} \simeq 17$\,kK and surface gravity $\log (g/{\rm cm\,s}^{-2}) \simeq 1.9$. To compute the extinction and luminosity, we assume the $VI$ photometry from August 19, 2022 (the closest multi-band photometry to the epoch of spectroscopic observations) of $V$=14.305, and $I$=13.931. We also justify this assumption since the $I$ magnitude is very similar to the epoch of spectroscopic observations ($I$=13.927 on November 10, 2024). Computing a bolometric correction in $I$ of $-$1.76, and a model ($V-I$)$_{\textrm{model}}$ of $-0.15$, we arrive at an $A_I$ of 0.77, leading to log\,($L/L_{\odot}$) of 4.73\,dex. 

As an alternative means of computing the luminosity, we rescale the spectral energy distribution of the best fitting \textsc{Fastwind} model to the aforementioned $I$ magnitude, adopting a distance modulus to the LMC of 18.48 \citep{2019Natur.567..200P} and accounting for the $I$ band extinction given above. This rescaling results in log\,($L/L_{\odot}$) of 4.57\,dex. We therefore quote an average value of the luminosity of [ST92] 4-13 as $10^{4.6\pm0.15}\,L_\odot$, where the uncertainties capture the discrepancies between the two methods. The object is shown on a Hertzsprung-Russel diagram along with known LBVs from the literature in Fig.\,\ref{hrd}.

\begin{table}
\centering
\caption{Best fitting parameters and 1$\sigma$ uncertainties for [ST92] 4-13 determined by \textsc{Kiwi-GA}.}
\label{tab:best_fit_params}
\begin{tabular}
{l r@{}l}
\midrule\midrule
Parameter & \multicolumn{2}{c}{Value} \\
\midrule
 $T_{\rm eff}$ [kK] & $17.2$ & $^{+0.3}_{-0.6}$ \\[3pt]
 $\log g$ $\left[{\rm cm~s}^{-2}\right]$ & $1.9$ & $^{+0.1}_{-0.1}$ \\[3pt]
 $Y_{\rm He}$ $\left[N_{\rm He}/N_{\rm H}\right]$ & $0.15$ & $^{+0.02}_{-0.04}$ \\[3pt]
 $v~\sin i$ $\left[{\rm km~s}^{-1}\right]$ & $76$ & $^{+2}_{-13}$ \\[3pt]
 $\log \dot{M}$ $\left[M_\odot\,{\rm yr}^{-1}\right]$ & $-5.7$ & $^{+0.1}_{-0.1}$ \\[3pt]
 $v_{\infty}$ $\left[{\rm km~s}^{-1}\right]$ & $114$ & $^{+14}_{-28}$ \\[3pt]
 $\log(N_{\rm C}/N_{\rm H}) + 12$  & $9.0$ & $^{+0.0}_{-0.4}$ \\[3pt]
 $\log(N_{\rm N}/N_{\rm H}) + 12$  & $8.2$ & $^{+0.1}_{-2.2}$ \\[3pt]
 $\log(N_{\rm O}/N_{\rm H}) + 12$  & $8.7$ & $^{+0.2}_{-0.0}$ \\[3pt]
 $\log(N_{\rm Si}/N_{\rm H}) + 12$  & $6.8$ & $^{+0.0}_{-0.2}$ \\[3pt]
 $R$ $\left[R_\odot\right]$ & $21.8$ & $^{+0.4}_{-0.3}$ \\[3pt]
\bottomrule
\end{tabular}
\end{table}

\subsubsection{Classification and evolutionary status of [ST92] 4-13}

The spectroscopic analysis points to change in spectral type compared to the literature, from B1a \citep{1992A&AS...92..729S} to approximately B2.5 (a change in $T_{\rm eff}$ of around 5kK based on the scale of \citealt{2007A&A...471..625T}). However, the derived luminosity is nearly an order of magnitude lower than that of the population of known LBVs \cite{smith04}. The star does show infrared \ion{Ca}{II} triplet seen in LBVs (but not the forbidden [\ion{Ca}{II}] emission, or any forbidden lines seen in B[e] stars as discussed by \citealt{hump1}), but not the classic P\,Cygni profiles of LBVs in this stage. The object is also not detected at longer wavelengths, ruling out a hot post-asymptotic giant branch (pAGB) classification (for e.g. see the discussion of the potential hot pAGB/LBV star WRA 571 in \citealt{wra571}), or even a B[e] classification (which have hot dust discs in the LMC, for e.g. \citealt{2019Galax...7...83K, 2014A&A...564L...7K}). 
Furthermore, no object currently known except LBVs exhibits these S\,Dor variations. But, based on its low luminosity and lack of P\,Cygni profiles in its spectra, we consider its status as a LBV ambiguous.

\begin{figure}
\center
\includegraphics[width=0.45\textwidth]{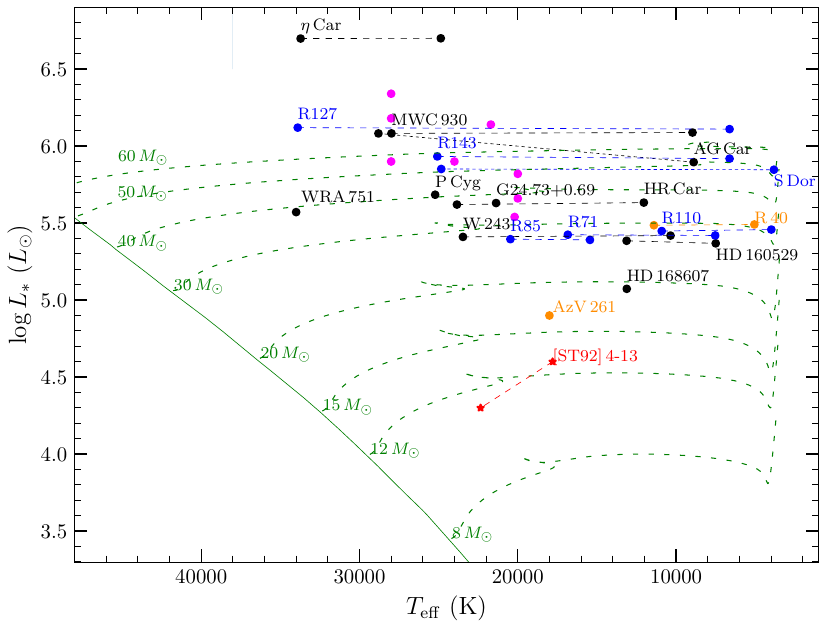}
\caption{ The location in the Hertzsprung-Russell diagram of [ST92]\,4-13 based on \cite{1992A&AS...92..729S} (left), and current observations (right) are shown in red. The positions of confirmed LBVs from the literature in the galaxy (black circles), LMC (blue circles), and the SMC (orange circle; including AzV\,261 discussed in Section 5.1) are also shown. Magenta circles marks the position of known LBVs in M31 and M33 (from \citealt{hump2}). The dashed green lines are the stellar tracks from Brott et al. (2011) with the corresponding initial stellar mass indicated, and the solid line is the zero-age main sequence isochrone for LMC metallicity. }
\label{hrd}
\end{figure}

\subsection{HD\,269216: Photometric and spectroscopic variability}

HD\,269216 was first suggested as a LBV candidate on the basis of ultraviolet and optical spectra by \cite{prinja91}. \cite{massey17} compiled archival photometric and spectroscopic information detailing the outburst between 2014 and 2016, where the spectrum changed from a late-B supergiant towards an A-F type supergiant. We concur with this analysis based on the {\it Gaia} light curve, the object has brightened considerably during the 2014-2016 outburst and is continuing that trend. Further monitoring of this object can track if it has brightened since late 2020.

\subsection{Circumstellar dust}

LBV variables show evidence of excess free-free emission \citep{hump1,hump2}, similar to classical Be stars, and not containing a warm dust component as found in other LBV-like objects such as B[e] stars. In Fig.\ref{jhk}, we show the near-infrared ($J-H$) versus ($H-K$s) color-color diagram showing the position of known massive stars, and also [ST92]\,4-13, and HD\,269216. In addition, we also show the data in the mid-infrared from WISE. In these diagrams, [ST92]\,4-13 exhibits modest free-free emission at levels similar to other known S\,Dor LBVs. HD\,269216 however resembles most main-sequence massive stars. Both resemble the colors of known LBVs \citep{hump1}.

\begin{figure*}
\center
\includegraphics[width=0.45\textwidth]{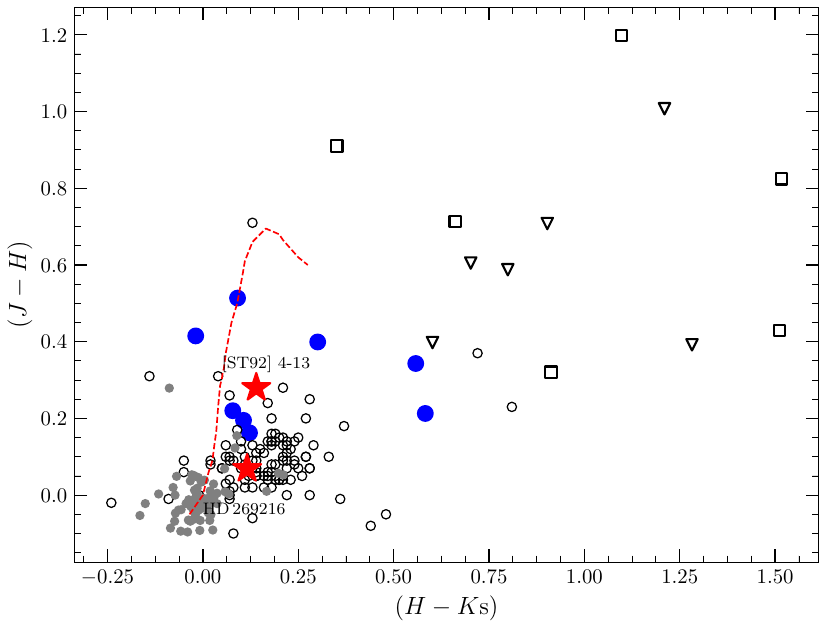}
\includegraphics[width=0.45\textwidth]{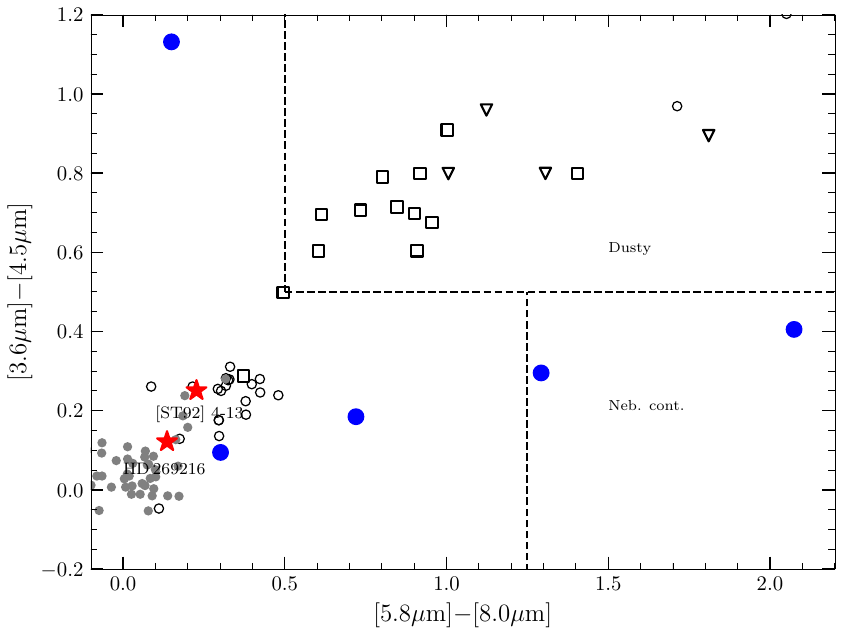} 
\caption{
{\it Left}: Near-infrared ($J-H$) vs. ($H-K$s) color-color diagram of known sgB[e] (open squares), warm hypergiants (open triangles), cBe stars (open circles), known S Dor variables (blue dots), BSgs 
(small dots), and [ST92]\,4-13, HD\,269216 (red asterisks). Data taken from \cite{hump1}. The dashed red line is the main-sequence locus. {\it Right}: Mid-infrared [3.6$\mu$m]$-$[4.5$\mu$m] vs. [5.8$\mu$m]$-$[8.0$\mu$m] color-color diagram.  
}
\label{jhk}
\end{figure*}

\section{Discussion} 

\subsection{S\,Dor variability and LBV classification}

In the literature, it is common practice to refer to massive stars with major (of which are only $\eta$\,Car, P\,Cyg) or minor outbursts (for e.g. S\,Dor) as LBVs. The latter are often characterized as S\,Dor variables, after it's namesake, but are often clumped together as bona fide LBVs. LBV candidates (LBVc), on the other hand are those having spectroscopic features (such as deep P\,Cygni profiles, spectroscopic transitions between late and early supergiants) but not the accompanying photometric variations at the S\,Dor level. 

In this paper, and in \cite{Kalari18}, we searched for evidence of S\,Dor variability among the BSg population in the LMC (and SMC in the earlier work). We found evidence of one candidate in the SMC which showed S\,Dor type variability without any spectral changes. Here, we show another candidate [ST92],4-13 in the LMC which demonstrates similar S\,Dor type variability over 20 years, but is likely not an LBV. 

This brings us to the questions: {\it are S\,Dor like photometric variations unique to LBVs?} Where do we classify S\,Dor like photometric variability without the accompanying spectral variations such as the objects discussed here. \cite{weiss, conti} have laid out that the defining characteristic of LBVs are the S\,Dor variability, or extremely rarely an $\eta$-Car like outburst. Here, we show evidence for two S\,Dor type variables, but with luminosity and spectral characteristics not resembling known LBVs.  

Potential hints may be drawn from the detection of known double-lined binaries, VFTS,698 and VFTS,652 as irregular variables in our analysis, although with amplitudes lower than the classical S,Dor variables. The former demonstrated long-term variability, but also B[e] spectrum (and infrared colors) and was classified as W-Serpentis variable, and consists of a massive binary around a cool primary. The latter was classified as a Algol-type variable, with a low-mass primary. Compared to these two, both AzV\,261 and [ST92]\,4-13 are single-lined (latter a binary based on multi-epoch radial velocity variations in the \ion{He}{i} line). It is possible that mass transfer across binary components may lead in specific cases to S\,Dor like light curves, such as seen in AzV\,261 \citep{2025A&A...698A..38B} or [ST92]\,4-13. Further high-resolution multi-epoch spectra of both these sources are necessary to understand the origin of these variations, and if they can be attributed to binary mass transfer, or if they are accompanied by spectral type changes. It may also be that most LBVs are born from mergers of hierarchical triple systems \citep{hirai}, and the two stars discussed here are lower-mass counterparts of such interactions. With the advent of multi-epoch photometry from surveys such as LSST, care must be taken in defining the LBV variability based on photometric variability. In this case, LBV should only be classified as such when observed with accompanying spectral characteristics. Drawing from the duck analogy laid out in \cite{conti} and \cite{2007AJ....134.2474M}, we suggest that if an LBV shows photometric and spectroscopic variability, it should be classified as an LBV, and if only one a LBVc. 

An alternative explanation for [ST92]\,4-13, and AzV\,261 could be that these objects are the observational indication of low-luminosity LBVs \citep{smith04}. It has been proposed using rotating evolutionary models \citep{2025ApJ...981..176L, lowlumlbv}, that stellar rotation lowers the mass (and thereby luminosity) threshold of the LBV phase. At solar metallicities, this is predicted to be around 22\,$M_{\odot}$, although the mass limit is thought to increase at lower metallicities, and is dependent on the rotational velocity.  

At LMC metallicity, for an initial velocity of 300\,$km\,s^{-1}$, \cite{lowlumlbv} find a lower initial mass limit of 25\,$M_{\odot}$. The initial mass of [ST92]\,4-13 is lower than this limit, hence we consider this possibility unlikely. Instead, we consider that the examples of the S\,Dor type phenomena at even lower luminosities (and by inference lower masses) reported here and in \cite{Kalari18} may not be linked to a single evolutionary phase, but rather are caused by Eddington-limit related physics at a range of masses, which may e.g. involve binary evolution.

\subsection{S\,Dor variables among the Bsg population}

The primary motivation of our study was to follow up on our previous results presented \cite{Kalari18}, but with a longer baseline and to cover a large sample size by studying BSgs in the Large Magellanic Cloud. In that study in the SMC, over a 3-year baseline of 64 BSgs we detected only one potential S\,Dor variable (AzV\,261), which did not show any spectral type variation. Here, covering a sample size of 87 (37) across a 20 (3) year timescale, we found two objects exhibiting S\,Dor type variations. One was the known LBV, HD\,269216, and the other a newly identified S\,Dor variable [ST92]\,4-13. The latter exhibits a potential change in spectral type from B1 (literature) to B2.5, but its luminosity is too low to be classified as an LBV. In addition, its spectra do not show the P\,Cygni profiles caused due to strong mass loss of LBVs. 

Under the assumption that all known LBVs have been discovered from the Bsgs population of the Large Magellanic Cloud (LMC) galaxy (around 8 (4) S\,Dor variables from 250 known Bsgs according to census compiled by \citealt{bonanos2009}), we estimate the duty cycle of the S\,Dor phase to be at most $\sim$few 10$^3$ yr (assuming a He-burning lifetime of 10$^5$ years). This is similar to the value found for the SMC \citep{Kalari18}. 

The work in the SMC was based on multi-epoch photometry covering a 3-year baseline. In this study, we expand the baseline and sample size. The similarity of the duty cycle estimated by the two studies at different metallicity values suggest that -- based on the current observational limitations (in magnitude range, and baselines) -- the S\,Dor duty cycle is at least an order of magnitude shorter than commonly assumed, and moreover BSgs and LBVs are truly different evolutionary stages in the life-cycle of massive stars. In other words, BSgs are not dormant LBVs.

\section{Conclusions}

\begin{itemize}
  \item We improved on previous searches for LBV variability among BSgs by increasing the baseline, and sample size from the previous work of \cite{Kalari18}. Here, we examined 20-year nearly nightly cadence light curves of 87 BSgs from OGLE, and 3 year {\it Gaia} multi-epoch photometry of 37 BSgs in the LMC. From examining these, we found evidence for S\,Dor type variability found in LBVs for one known LBV (HD\,269216), and one BSg ([ST92]\,4-13).   
  \item However, despite an increased sample size and baseline, the duty cycle we found for LBVs was roughly similar, around 10$^3$ years. This suggests that LBVs are truly unique or transitory objects, and that the LBV phenomenon is likely not widespread amongst the general BSg population. 
  \item We also argue that both photometric and spectral type variations, along with spectral characteristics and high-luminosities are necessary to classify LBVs, based on the properties of the candidates presented here. 
  \item  Our findings suggest that LBV-like variability is a physical phenomena, and not and not solely linked to any one particular evolutionary phase.
\end{itemize}

\begin{acknowledgements}
Table 1 is only available in electronic form at the CDS via anonymous ftp to cdsarc.u-strasbg.fr. This research has made use of the SIMBAD database, CDS, Strasbourg Astronomical Observatory, France. The OGLE project has received funding from the Polish National Science Centre grant OPUS-28 2024/55/B/ST9/00447 to AU. This work presents results from the European Space Agency (ESA) space mission Gaia. Gaia data are being processed by the Gaia Data Processing and Analysis Consortium (DPAC). Funding for the DPAC is provided by national institutions, in particular the institutions participating in the Gaia MultiLateral Agreement (MLA). The Gaia mission website is https://www.cosmos.esa.int/gaia. The Gaia archive website is https://archives.esac.esa.int/gaia. V.M.K. is supported by the international Gemini Observatory, a program of NSF NOIRLab, which is managed by the Association of Universities for Research in Astronomy (AURA) under a cooperative agreement with the U.S. National Science Foundation, on behalf of the Gemini partnership of Argentina, Brazil, Canada, Chile, the Republic of Korea, and the United States of America.

\end{acknowledgements}

%
%

\bibliographystyle{aa} 
\bibliography{38.bib} 

\begin{appendix}
    
\section*{Appendix A: $\delta m-\delta t$ histograms and $\eta$ index of confirmed LMC LBVs}
\label{appendix:lbvs}

We retrieved {\it Gaia} DR3 multi-epoch photometry of confirmed LBVs listed in \cite{2018RNAAS...2..121R} (see Section~2.3). There are 7 sources, all of which had multi-epoch photometry spanning the DR3 period (3 years). We show in Fig.\,\ref{fig:seven_panels} the histograms computed as described in Section 3.1, utilizing the $G$-band photometry. Only data points with {\texttt{variability\_flag\_g\_reject}}=False were used in any analysis. The histograms all have aperiodic variability above the significance level. The distribution of the histograms are similar to the LBV HD\,269216 and [ST92]\,4-13, but the histogram of [ST92]\,4-13 also shows a periodic trend along with aperiodic variability (both vertical and inclined lines).

In the top-right of each histogram, the 1/$\eta$ index is given, and computed as described in Section\,3.2. All stars have significant indexes except for HD\,269700, representative of aperiodic variability. The lack of variability is not surprising for HD\,269700, since it is considered an ex/dormant LBV \citep{1999A&A...349..537V}. All these stars would be selected as an LBV with our analysis methodology, indicating our method is suitable to recover LBVs. Future multi-epoch photometry from {\it Gaia}\footnote{https://www.cosmos.esa.int/web/gaia/dr4\#} covering a much larger baseline can be used to search for further possible LBVs in the Magellanic Clouds.

\begin{figure*}[p]
    \centering
    \begin{subfigure}{0.45\textwidth}
        \centering
        \includegraphics[width=\linewidth]{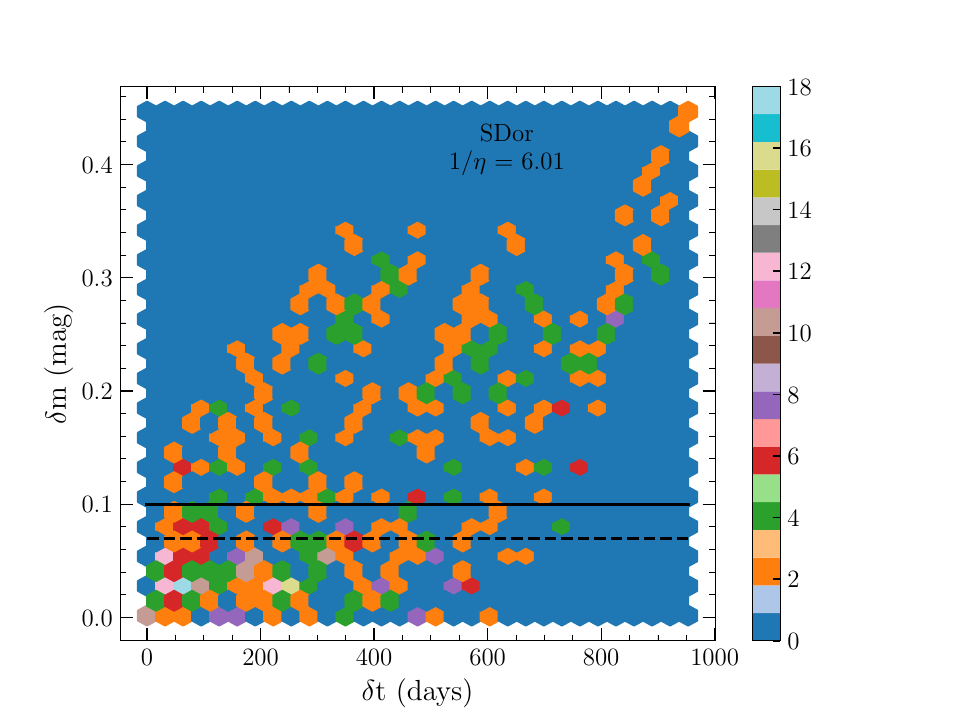}
    \end{subfigure}\hfill
    \begin{subfigure}{0.45\textwidth}
        \centering
        \includegraphics[width=\linewidth]{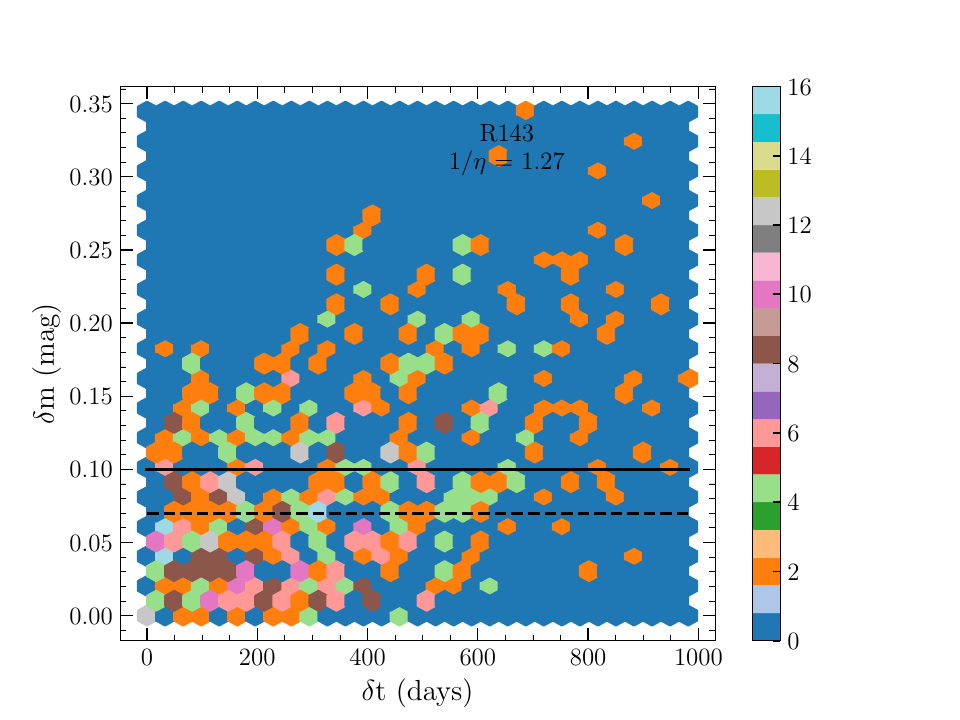}
    \end{subfigure}
    \begin{subfigure}{0.45\textwidth}
        \centering
        \includegraphics[width=\linewidth]{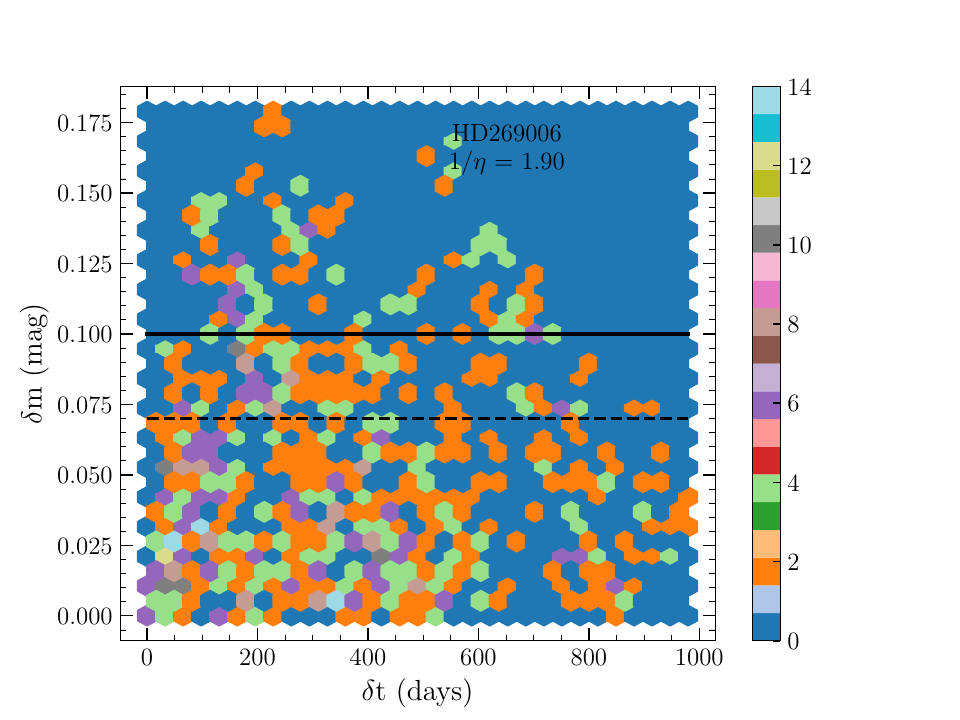}
    \end{subfigure}\hfill
    \begin{subfigure}{0.45\textwidth}
        \centering
        \includegraphics[width=\linewidth]{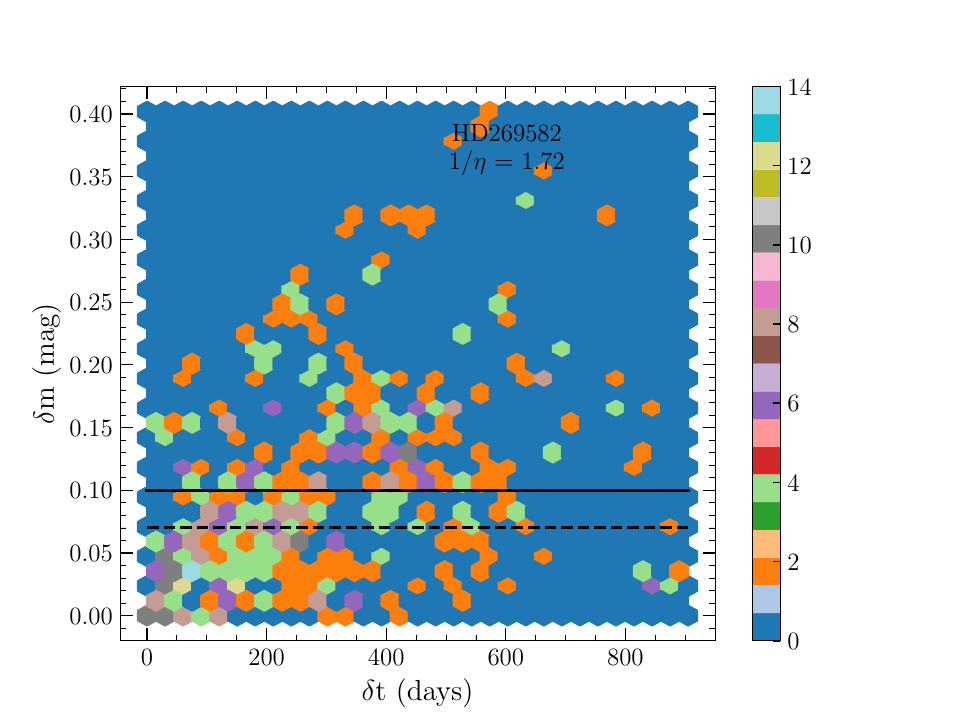}
    \end{subfigure}
    \begin{subfigure}{0.45\textwidth}
        \centering
        \includegraphics[width=\linewidth]{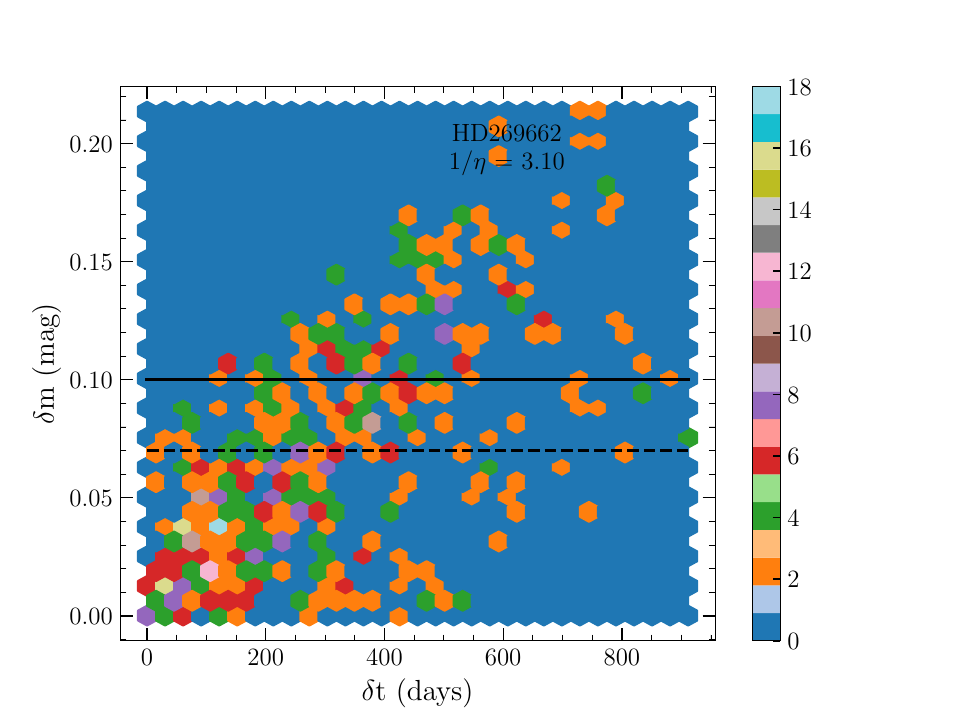}
    \end{subfigure}\hfill
    \begin{subfigure}{0.45\textwidth}
        \centering
        \includegraphics[width=\linewidth]{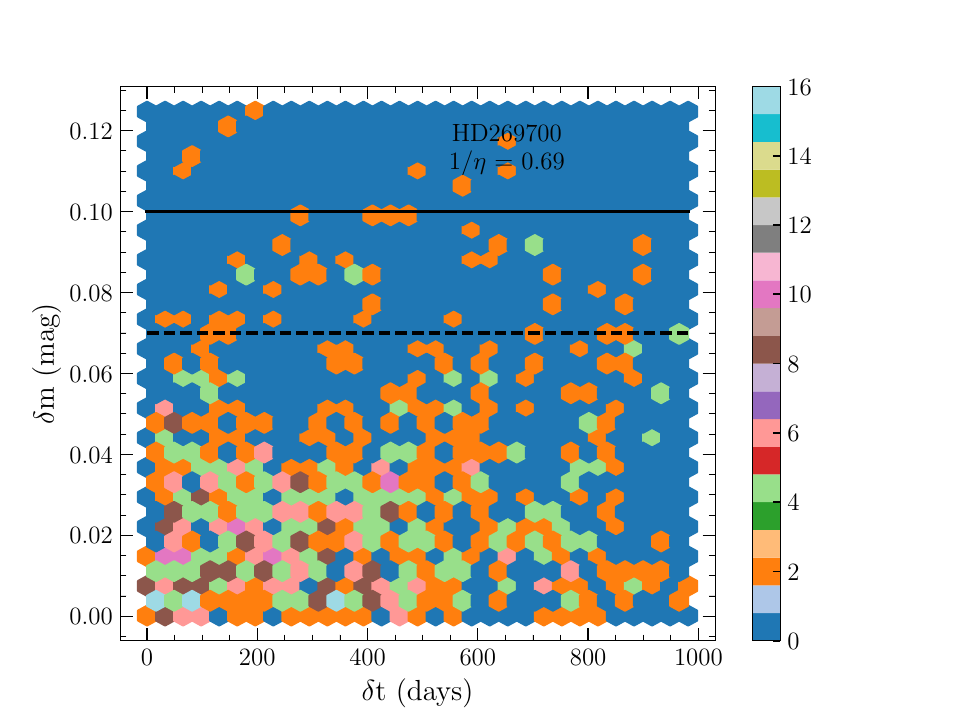}
    \end{subfigure}
    \begin{subfigure}{0.45\textwidth}
        \centering
        \includegraphics[width=\linewidth]{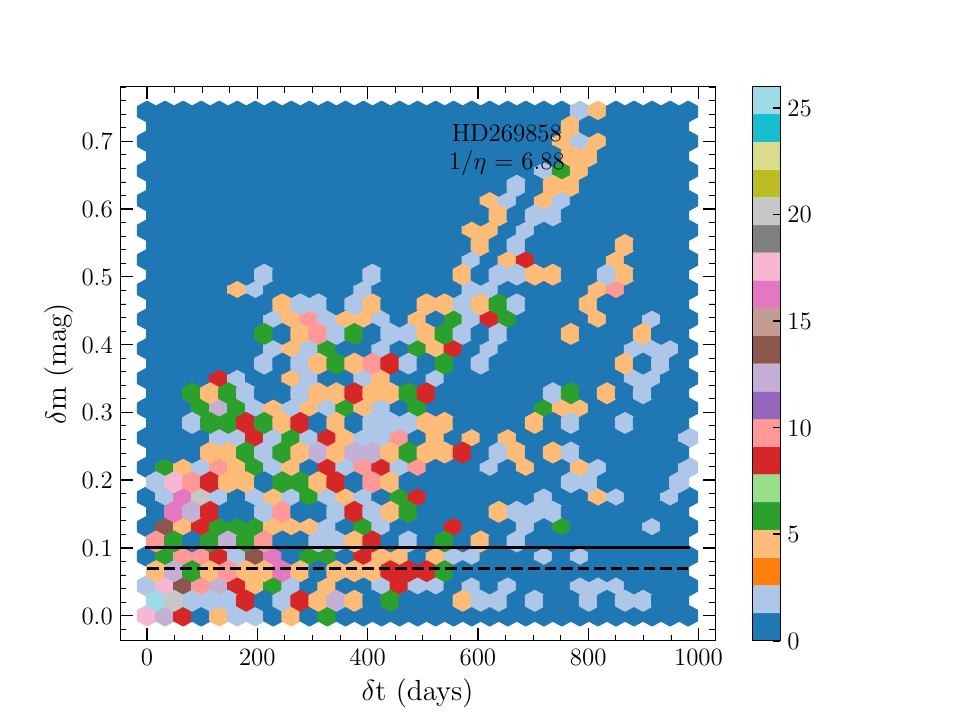}
    \end{subfigure}
    \caption{$\delta m-\delta t$ histograms of confirmed LMC LBVs, with the 1/$\eta$ value given in the top-right.}
    \label{fig:seven_panels}
\end{figure*}

\section*{Appendix B: Modeling approach for [ST92] 4-13}
\label{appendix:modelling_approach}

In this section, we describe how we prepared the optical spectrum of [ST92] 4-13 to be analyzed. 

First, we decided on a list of diagnostic spectral lines. With the main goal of this exercise being to gauge the evolutionary status of our target, we are primarily interested in the determination of the effective temperature ($T_{\rm eff}$) and surface gravity ($g$) from its spectrum, and the luminosity ($L$) from these. For $T_{\rm eff}$, we used the \ion{Si}{ii} to \ion{Si}{iii} ratio; the primary temperature diagnostic in B stars \citep[e.g.][]{1999A&A...349..553M}. We used the photospheric \ion{He}{i} lines to constrain $g$. As these lines show wind features, we also use them to get an idea of the mass loss rate ($\dot{M}$) and terminal wind speed ($v_{\infty}$). Regarding the wind structure, we assumed a fixed clumping factor of 10 and that all clumps in the wind are optically thin. We also estimate the projected rotational velocity, $v\,\sin\,i$. We do not account for macroturbulent broadening here, so our quoted value of $76^{+2}_{-13}$\,km\,s$^{-1}$ in Table \ref{tab:best_fit_params} is an upper limit of the true projected rotational velocity. Finally, we left the helium abundance ($Y_{\rm He} := N_{\rm He} /N_{\rm H}$), and C, N, O, and Si abundances as free parameters in our fit.

While the spectrum of our target exhibits strong Balmer emission, indicative of a dense stellar wind, we do not fit these lines here. Our current assumptions regarding the wind structure are not enough to adequately describe the nature of the stellar wind; simultaneous modelling of the UV spectrum, where resonance lines of metals lie, would be needed in order to gain a complete picture of the wind of this star. From a test run where the Balmer lines were included, we found they strongly biased $T_{\rm eff}$ and $g$ to very high values in order to compensate for our assumptions regarding the wind structure, resulting in an overall very poor fit. In any case, we are not interested in $\dot{M}$ here anyway, so for these reasons we left them out.

\textsc{Fastwind} and \textsc{Kiwi-GA} requires a radial velocity corrected, continuum normalised spectrum as input. To correct for radial velocity, we cross correlate with a template of Gaussian absorption profiles at the wavelengths of the photospheric \ion{He}{I} profiles, as well as at the three components of the \ion{Si}{iii}\,$\lambda 4452$ profile. From this we find a radial velocity of $269 \pm 38$\,km\,s$^{-1}$.

Once corrected for radial velocity, we locally normalised the target spectral lines. For a given line, this involved masking spectral features blue and red of the line, fitting a line through the local continuum level, and dividing through by this line. This is a reasonable approach given the relatively few spectral lines in the optical spectrum. In some instances, it was necessary to fit degree 2 polynomials to the continuum. This was the case for \ion{C}{iii}\,$\lambda 4070$, \ion{He}{i}\,$\lambda 4143$, and the \ion{O}{ii} doublet at $4416$\,\AA.

\end{appendix}

\end{document}